\documentclass{article}

\usepackage{microtype}
\usepackage{graphicx}
\usepackage{subfigure}
\usepackage{booktabs} 

\usepackage{hyperref}
\usepackage{enumitem}

\usepackage{amsmath}
\usepackage{amssymb}
\usepackage{mathtools}
\usepackage{amsthm}

\usepackage[accepted]{ml4astro2025}
\usepackage{float}
\usepackage{tikz}
\usepackage[capitalize,noabbrev]{cleveref}
\usetikzlibrary{positioning,shapes.multipart,calc,shadows,arrows.meta}
\usetikzlibrary{decorations.pathreplacing}
\usetikzlibrary{calc}
\tikzset{
  basic/.style={draw, text centered},
  circ/.style={basic, circle, minimum size=2em, inner sep=1.5pt},
  rect/.style={basic, text width=1.5em, text height=1em, text depth=.5em},
  1 up 1 down/.style={basic, text width=1.5em, rectangle split, rectangle split horizontal=false, rectangle split parts=2},
}
\theoremstyle{plain}

\theoremstyle{definition}

\theoremstyle{remark}

\usepackage[textsize=tiny]{todonotes}

\mlforastrotitlerunning{IrisML}

\begin{document}

\twocolumn[
\mlforastrotitle{IrisML: Neural Posterior Estimation for the Spectral Energy Distribution fitting}

\begin{mlforastroauthorlist}
    \mlforastroauthor{Mateusz Kapusta}{warsaw}
\end{mlforastroauthorlist}

\mlforastroaffiliation{warsaw}{Astronomical Observatory, University of Warsaw, Warsaw, Poland}

\mlforastrocorrespondingauthor{Mateusz Kapusta}{mr.kapusta@student.uw.edu.pl} 

\vskip 0.3in
]
\printAffiliationsAndNotice{}

\begin{abstract}
    Over the past 30 years, numerous large-scale photometric astronomical surveys have been conducted,
    including SDSS, Pan-STARRS, Gaia, 2MASS, WISE, and others. These surveys provide extensive
    photometric measurements that can be used to infer a wide range of physical parameters of astronomical objects.
    Traditionally, Bayesian approaches, such as Markov Chain Monte Carlo (MCMC) sampling—have been employed for such inference tasks.
    However, these methods tend to be computationally intensive and often require manual tuning or expert supervision.
    In this work, we propose a novel machine learning model designed to perform automatic and robust inference from photometric data,
    offering a scalable and efficient alternative to conventional techniques.
\end{abstract}

\section{Introduction}
For decades, astronomical research has relied on broadband photometry to uncover the physical parameters of celestial objects.
Since the advent of space-based observations, the volume of data available to researchers has been growing rapidly,
setting high expectations for data preprocessing pipelines. Various astronomical surveys 
like the upcoming LSST are producing terabytes of data across different parts of the optical spectrum, enabling the estimation of parameters for millions of stars.

Using theoretical spectral models, broadband photometry can be employed to estimate stellar parameters
through Spectral Energy Distribution (SED) fitting. Typically,
Bayesian methods such as Markov Chain Monte Carlo (MCMC) and Variational Inference are used to infer parameters in such complex models.
Notable examples include research on interstellar extinction \citep{green_3d_2019, leike_charting_2019},
which demands substantial computational power for full Bayesian inference.
Other key applications of SED fitting include detecting secondary stellar components \citep{el-badry_what_2022},
estimating redshifts from photometry \citep{william_deep_2023}, and determining galaxy properties \citep{pacifici_art_2023}.

Recent advances in computational methods, particularly the emergence of neural network-based approaches,
have led to a new class of algorithms designed to accelerate inference in various astronomical tasks.
Examples include X-ray spectral fitting \citep{huppenkothen_accurate_2022}, gravitational wave astronomy \citep{dax_real-time_2021},
microlensing \citep{zhang__real-time_2021}, and the inference of galactic properties \citep{li_popsed_2024}.
These models often rely on techniques from Neural Density Estimation (NDE), which fall under the broader category of Likelihood-Free Inference (LFI).
NDE methods offer significant advantages over traditional MCMC approaches, enabling faster and more automated inference.
This makes them particularly well-suited for real-time data processing pipelines.

Here, we present a new NDE-based model that enables fast inference of stellar parameters such as temperature, extinction, metallicity,
and surface gravity from broadband photometry. In contrast to previous approaches, which rely on a fixed set of filters for inference,
the proposed model supports variable-sized input. This flexibility allows for the seamless inclusion of new filters
and updating of the model without the need to retrain it from scratch.

\section{Inference Task}
The primary objective of the model is to infer three fundamental stellar parameters using broadband photometric measurements:
the effective temperature $T$, surface gravity $\log g$, and metallicity [M/H]. In addition, the model estimates two dust-related parameters:
extinction in the $V$ band, denoted as $A_V$, and the total-to-selective extinction ratio, $R_V$.

One of the most challenging aspects of the modeling process lies in the efficient incorporation of an arbitrary number of photometric measurements.
This step is critical for accurately estimating stellar parameters in complex regions of the sky, such as the Galactic bulge.
In such areas, high levels of extinction can make stars appear significantly cooler than they actually are.
Resolving this degeneracy requires additional observations, particularly in the far-infrared regime.
Restricting the model to a fixed set of filters limits its effectiveness: stars with missing data
in certain bands require special handling, while those with more comprehensive observations cannot fully exploit the additional information.
Addressing this limitation necessitates the use of more sophisticated and flexible model architectures.

Photometric data are inherently less structured than many other types of datasets typically encountered in machine learning.
Each individual measurement consists of three components:
\begin{itemize}[itemsep=0pt, parsep=0pt, topsep=0pt, partopsep=0pt]
    \item the photometric filter used,
    \item the observed brightness (in magnitudes),
    \item the associated measurement uncertainty (also in magnitudes).
\end{itemize}
A robust model must be capable of integrating these diverse elements while accommodating a variable number of measurements per object.

\section{Methods}
Proposed model is composed of two different parts: set-invariant transformer and Masked Autoregressive Flow (MAF) module \citep{papamakarios_masked_2017}.
The first submodel is responsible for the preprocessing step, it allows to take any number of filters and compress the information to a single vector.
Then, compressed vector is used to feed the MAF model.

\subsection{Preprocessing}
The preprocessing model is built upon a set-invariant transformer architecture. Suppose there are $S$ different photometric measurements for a given star. Each measurement is converted into an $M$-dimensional token, which serves as input to the transformer.

To construct each token:
\begin{itemize}[itemsep=0pt, parsep=0pt, topsep=0pt, partopsep=0pt]
    \item The photometric filter is encoded using a learned embedding.
    \item The brightness value is processed through a linear layer to generate an embedding.
    \item The associated measurement uncertainty is processed through a linear layer to generate an embedded.
\end{itemize}

These components are summed to form the final token representation for each observation.

To aggregate the $S$ observation tokens, we employ a
classification token—a technique originally introduced in \citet{BERT}
for sentence classification using transformers. In our context, this token is used to
compress the set of measurements into a single, fixed-size representation.

The transformer receives a sequence consisting of this learnable classification token,
followed by the $S$ embedded observation tokens. This sequence is passed through $L$ layers of
multi-head self-attention and feed-forward networks.
After processing, the updated classification token encodes the compressed representation of the entire set.
This token is then passed to a Masked Autoregressive Flow (MAF) model as a conditional input for downstream inference.

Thanks to the transformer's permutation-invariant design, the final output is independent of the order in which the measurements are presented.
For our implementation:
\begin{itemize}[itemsep=0pt, parsep=0pt, topsep=0pt, partopsep=0pt]
    \item The embedding dimension $M$ is set to $256$.
    \item We use $L = 10$ transformer layers.
    \item Each feed-forward network has a dimensionality of $768$.
    \item The model uses $8$ attention heads.
    \item Dropout with a rate of $p = 0.05$ is applied during training.
\end{itemize}

\subsection{Masked Autoregressive Flow}
The Masked Autoregressive Flow is a model belonging to the family of Normalizing Flows,
a class of generative models originally developed for tasks such as high-fidelity image synthesis,
with early breakthroughs seen in models like GLOW \citep{GLOW}.

The core concept of normalizing flows is straightforward yet powerful.
These models aim to estimate the conditional probability distribution $P(X \mid C)$,
where $C$ represents the conditioning variable (in this context, stellar observations),
and $X$ denotes the target variables (e.g., stellar parameters such as temperature, surface gravity, and metallicity). 

To achieve this, a sequence of invertible transformations is learned such that the final function 
$f$ transforms the conditional distribution $P(X|C)$ into the base distribution $\pi(u)$. This base distribution is commonly a multivariate Gaussian—chosen
for its ease of evaluation and sampling.

The MAF architecture is composed of a series of MADE (Masked Autoencoder for Distribution Estimation)
sub-blocks \citep{germain_made_2015}, where each block implements a single transformation $f_i$ of the
probability space. These transformations are composed sequentially as $f = f_1 \circ f_2 \circ f_3 \circ \ldots$,
enabling the model to approximate highly complex mappings. For brevity, the detailed mechanics of MAF are omitted here,
as the architecture used closely follows implementations extensively described in the literature.

During inference, sampling proceeds by first drawing $u$ from the base distribution $\pi(u)$
and then applying the inverse transformation $f^{-1}$ to obtain a sample from the learned posterior.
This approach allows for extremely fast approximation of complex posterior distributions.

\begin{figure*}
    \begin{tikzpicture}[scale=0.25, every node/.style={scale=2.5},transform shape]
        \node[draw,circle,fill=pink] at (0,0) (class) {CLS};
        \node[rectangle,rounded corners,thick, fill=blue!20,]  at (5,0) {Token 1};
        \node[rectangle,rounded corners,thick, fill=blue!20,]  at (10,0) {Token 2};
        \node[rectangle,rounded corners,thick, fill=blue!20,]  at (15,0) {Token 3};
        \node[text width=1.5cm] at (20,0) {$\ldots$};
        \node[rectangle,rounded corners,thick, fill=blue!20,]  at (23,0) {Token $S$};

        \draw[->,black] (0,1.1) to (0,1.8);
        \draw[->,black] (5,0.6) to (5,1.8);
        \draw[->,black] (10,0.6) to (10,1.8);
        \draw[->,black] (15,0.6) to (15,1.8);
        \draw[->,black] (23,0.6) to (23,1.8);

        \draw (-2,2) rectangle node{Multi Head Attention} (27,4);
        \draw (-2,4) rectangle node{Feed Forward} (27,6);
        \node[text width=1.5cm] at (14,8) {$\vdots$};
        \draw (-2,10) rectangle node{Multi Head Attention} (27,12);
        \draw (-2,12) rectangle node{Feed Forward} (27,14);
        \node[draw,circle,fill=pink] at (0,16) (class_out) {CLS};
        \node[rectangle,rounded corners,thick, fill=blue!20,] at (5,16) {Token 1};
        \node[rectangle,rounded corners,thick, fill=blue!20,] at (10,16) {Token 2};
        \node[rectangle,rounded corners,thick, fill=blue!20,] at (15,16) {Token 3};
        \node[text width=1.5cm] at (20,16) {$\ldots$};
        \node[rectangle,rounded corners,thick, fill=blue!20,] at (23,16) {Token $S$};

        \draw[->,black] (0,14.2) to (0,14.8);
        \draw[->,black] (5,14.2) to (5,15.3);
        \draw[->,black] (10,14.2) to (10,15.3);
        \draw[->,black] (15,14.2) to (15,15.3);
        \draw[->,black] (23,14.2) to (23,15.3);

        \node[rectangle,rounded corners,thick, fill=green!20,] at (0,-8) {Filter 1};
        \node[rectangle,rounded corners,thick, fill=green!20,] at (5,-8) {Magnitude 1};
        \node[rectangle,rounded corners,thick, fill=green!20,] at (10,-8) {Error 1};

        \node[rectangle,rounded corners,thick, fill=green!20,] at (0,-5) {Learned embedding};
        \node[rectangle,rounded corners,thick, fill=green!20,] at (5,-5) {Linear};
        \node[rectangle,rounded corners,thick, fill=green!20,] at (10,-5) {Linear};
        \draw[->,black] (0,-7.2) to (0,-5.5);
        \draw[->,black] (5,-7.2) to (5,-5.5);
        \draw[->,black] (10,-7.2) to (10,-5.5);

        \draw[->,black] (0,-4.5) to[in = -135] (4,-2);
        \draw[->,black] (5,-4.5) to (5,-2.8);
        \draw[->,black] (10,-4.5) to[out=135,in = -45] (6,-2);

        \node[draw, circle] at (5,-2){$+$};
        \draw[->,black] (5,-1.3) to (5,-0.5);

        \draw [decorate,decoration={brace,amplitude=5pt,mirror,raise=4ex}]
                (25,2) -- (25,14) node[midway,xshift=5em]{$L$};
        \coordinate (Center_MADE) at (36,6);
        \node[rectangle,rounded corners,thick, fill=red!20,] at ($ (Center_MADE) + (-5,0)$) (Y) {$P(X|C)$};
        \node[rectangle,rounded corners,thick, fill=red!20,] at ($ (Center_MADE) + (29,0)$) (U) {$\pi(u)$};
        \node[rectangle,rounded corners,thick, fill=red!20,] at ($ (Center_MADE)$) (M1) {MADE 1};
        \node[rectangle,rounded corners,thick, fill=red!20,] at ($ (Center_MADE) + (8,0) $) (M2) {MADE 2};
        \node[rectangle,rounded corners,thick, fill=red!20,] at ($ (Center_MADE) + (15.5,0)$) (M0) {$\ldots$};
        \node[rectangle,rounded corners,thick, fill=red!20,] at ($ (Center_MADE) + (23,0)$)  (MM) {MADE $N$};
        \draw[-] (class_out) to ( $(class_out) + (0,4)$);
        \draw[-] ( $(class_out) + (0,4)$) to ( $(class_out) + (30,4)$);
        \draw[->] ( $(class_out) + (30,4)$)to[out=0,in=90] (M1);
        \draw[->] ( $(class_out) + (30,4)$) to[out=0,in=90] (M2);
        \draw[->] ( $(class_out) + (30,4)$) to[out=0,in=90] (M0);
        \draw[->] ( $(class_out) + (30,4)$) to[out=0,in=90] (MM);
        \draw [->,red,thick] (M1) to [out=20,in=160] (M2);
        \draw [->,orange,thick] (M2) to [out=-160,in=-20] (M1);

        \draw [->,red,thick] (M2) to [out=20,in=160] (M0);
        \draw [->,orange,thick] (M0) to [out=-160,in=-20] (M2);

        \draw [->,red,thick] (M0) to [out=20,in=160] (MM);
        \draw [->,orange,thick] (MM) to [out=-160,in=-20] (M0);

        \draw [->,red,thick] (Y) to [out=20,in=160] (M1);
        \draw [->,orange,thick] (M1) to [out=-160,in=-20] (Y);

        \draw [->,red,thick] (MM) to [out=20,in=160] (U);
        \draw [->,orange,thick] (U) to [out=-160,in=-20] (MM);

        \draw [->,red,thick] ($(Center_MADE) + (8,-3)$) to ($(Center_MADE) + (10,-3) $);
        \node[text width=3cm] at ($(Center_MADE) + (14,-3)$) {training};

        \draw [->,orange,thick] ($(Center_MADE) + (8,-4)$) to ($(Center_MADE) + (10,-4) $);
        \node[text width=3cm] at ($(Center_MADE) + (14,-4)$) {inference};

    \end{tikzpicture}
    \caption{Schematic representation of the IrisML model. CLS stands for the classification token, which is learned during the training. Red and orange 
    arrows indicate the movement of the data during the training and inference respectively.
    }
    \label{tikz:flow}
\end{figure*}
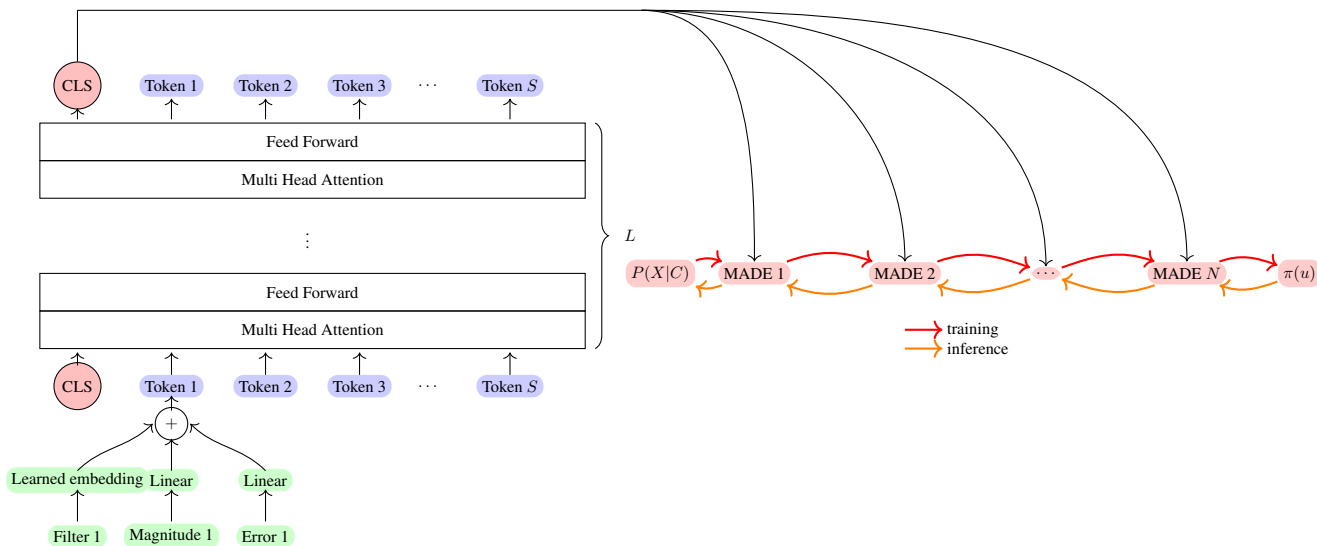

Following the methodology of \citep{zhang__real-time_2021}, we adopt a multivariate Gaussian base distribution with five components.
The flow consists of ten stacked MADE blocks. The dimensionality of the conditioning input $C$ corresponds to the token size used in our encoder,
which in this case is $256$. An overview of the final architecture is provided in Figure \ref{tikz:flow}.

\subsection{Training and data generation}
The training data for the model are derived from the BOSZ spectral library \citep{BOSZ},
combined with the Fitzpatrick extinction law \citep{fitzpatrick_correcting_1999}.
To interpolate stellar spectra, the \texttt{pystellibs}\footnote{\url{https://mfouesneau.github.io/pystellibs/}} library was employed,
while extinction effects were modeled using the \texttt{extinction}\footnote{\url{https://extinction.readthedocs.io/en/latest/}} package.

A total of $25$ photometric filters were incorporated into the model,
spanning from the far-ultraviolet to the infrared. The complete list of filters,
along with their respective survey catalogs to be utilized in the later stages of this study, includes:
\begin{itemize}[itemsep=0pt, parsep=0pt, topsep=0pt, partopsep=0pt]
    \item \textbf{GALEX} ultraviolet filters: FUV, NUV \citep{martin_galaxy_2005}
    \item \textbf{SDSS} optical filters: $u$, $g$, $r$, $i$, $z$ \citep{kollmeier_sdss-v_2019}
    \item \textbf{Pan-STARRS} optical filters: $g$, $r$, $i$, $z$, $y$ \citep{chambers_pan-starrs1_2016}
    \item \textbf{SkyMapper} optical filters: $g$, $r$, $i$, $z$, $y$ \citep{australian_national_university_skymapper_2023}
    \item \textbf{Gaia} optical filters: $G$, $BP$, $RP$ \citep{gaia_collaboration_gaia_2023}
    \item \textbf{2MASS} infrared filters: $J$, $H$, $K_s$ \citep{skrutskie_two_2006}
    \item \textbf{WISE} infrared filters: $W1$, $W2$ \citep{wright_wide-field_2010}
\end{itemize}

In total, approximately $1.5 \times 10^6$ synthetic stars were generated for training,
and about $3 \times 10^5$ for testing purposes. During training, each photometric measurement was
dynamically perturbed with Gaussian noise, simulating measurement uncertainty. This noise was passed as a variable to the preprocessing pipeline.

The training process was conducted over $1000$ epochs using the Adam optimizer \citep{kingma_adam_2014}.
The learning rate followed a cosine annealing schedule with a linear warm-up phase, starting from a base value of $5 \times 10^{-4}$.


\section{Results}
To comprehensively evaluate the final performance of our model, we design two distinct validation tasks.

First, we assess the model’s ability to predict stellar temperatures in highly reddened regions.
Accurate temperature estimation in such environments is essential for reliable reddening corrections,
which in turn are a key component in constructing three-dimensional maps of interstellar dust.

Second, we evaluate the accuracy of the model in predicting stellar metallicities.
This parameter is particularly critical for low-metallicity stars,
which have been the focus of extensive research over the past decade.

To validate temperature predictions, we select a sample of stars from the APOGEE survey \citep{abdurrouf_seventeenth_2022},
focusing on objects located close to the Galactic plane ($b < 20^\circ$), hereafter referred to as the \textit{Disc sample}.
Each star is cross-matched with multiple external catalogs to ensure broad observational coverage,
and only stars with more than 10 independent photometric observations are retained. The APOGEE
survey fits the spectra using precomputed high-resolution spectra. APOGEE errors are roughly of the order of $\sim 150$ K. 

\begin{figure}
    \includegraphics[width = 0.5\textwidth]{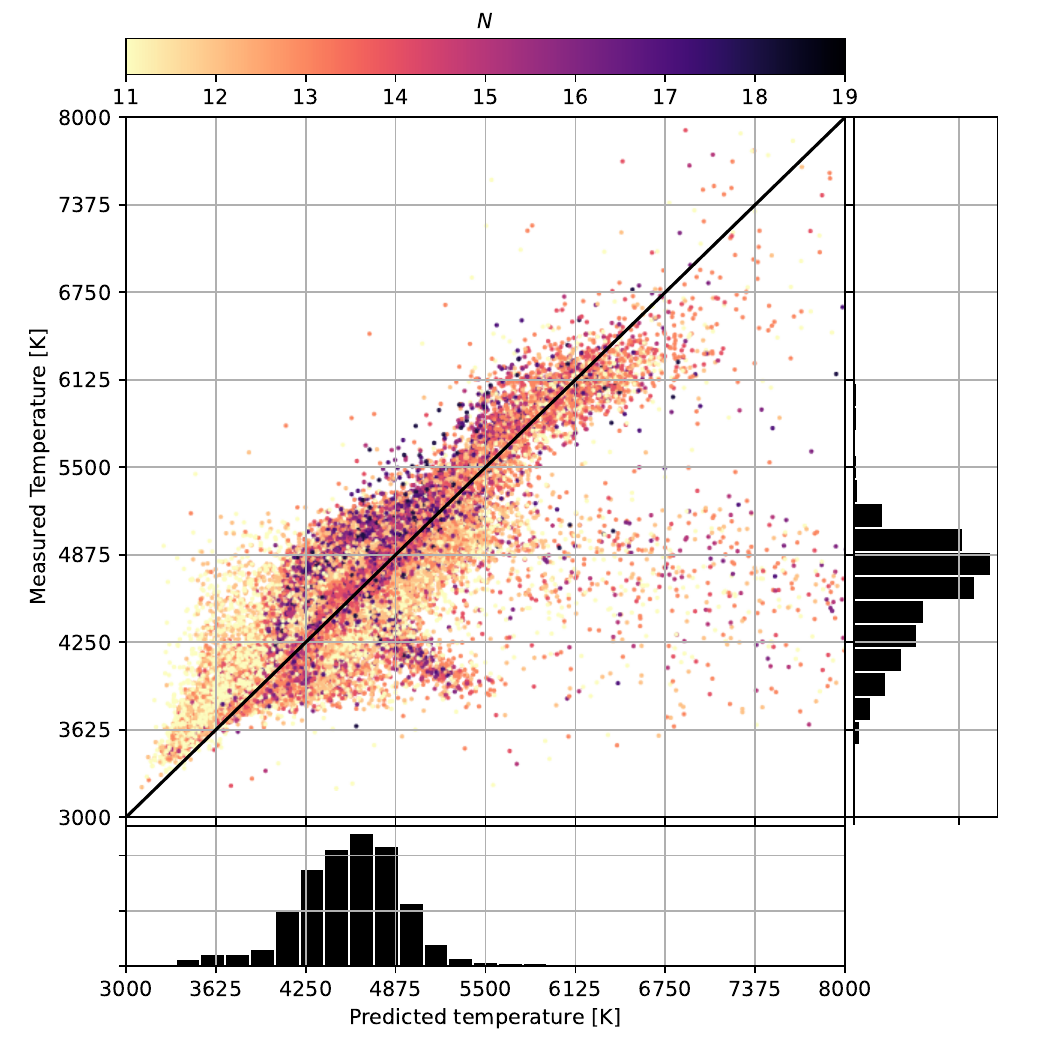}
    \caption{Predicted and measured temperatures for stars in the Disc sample. Color represents the number of measurments used to infer the stellar properties.}
    \label{fig:disc}
\end{figure}

We then apply our model to estimate stellar temperatures, using the median of 64 posterior samples as the final prediction.
These are compared against APOGEE-derived temperatures. Across a total of 124,428 stars,
the model achieves a root-mean-square (RMS) error of $314$ K and a median absolute deviation (MAD) of $169$ K, even in the presence of significant reddening.
The comparison between predicted and true temperature is presented in Figure \ref{fig:disc}.

Although APOGEE does not provide direct extinction measurements, the strong correlation between temperature and
extinction allows us to infer that extinction estimates are likely to be reasonably accurate.
Mean predicted extinction in the $V$ filter $A_V$ is $1.93$, with a $84$th percentile reaching $3.42$.

\begin{figure}
    \includegraphics[width = 0.5\textwidth]{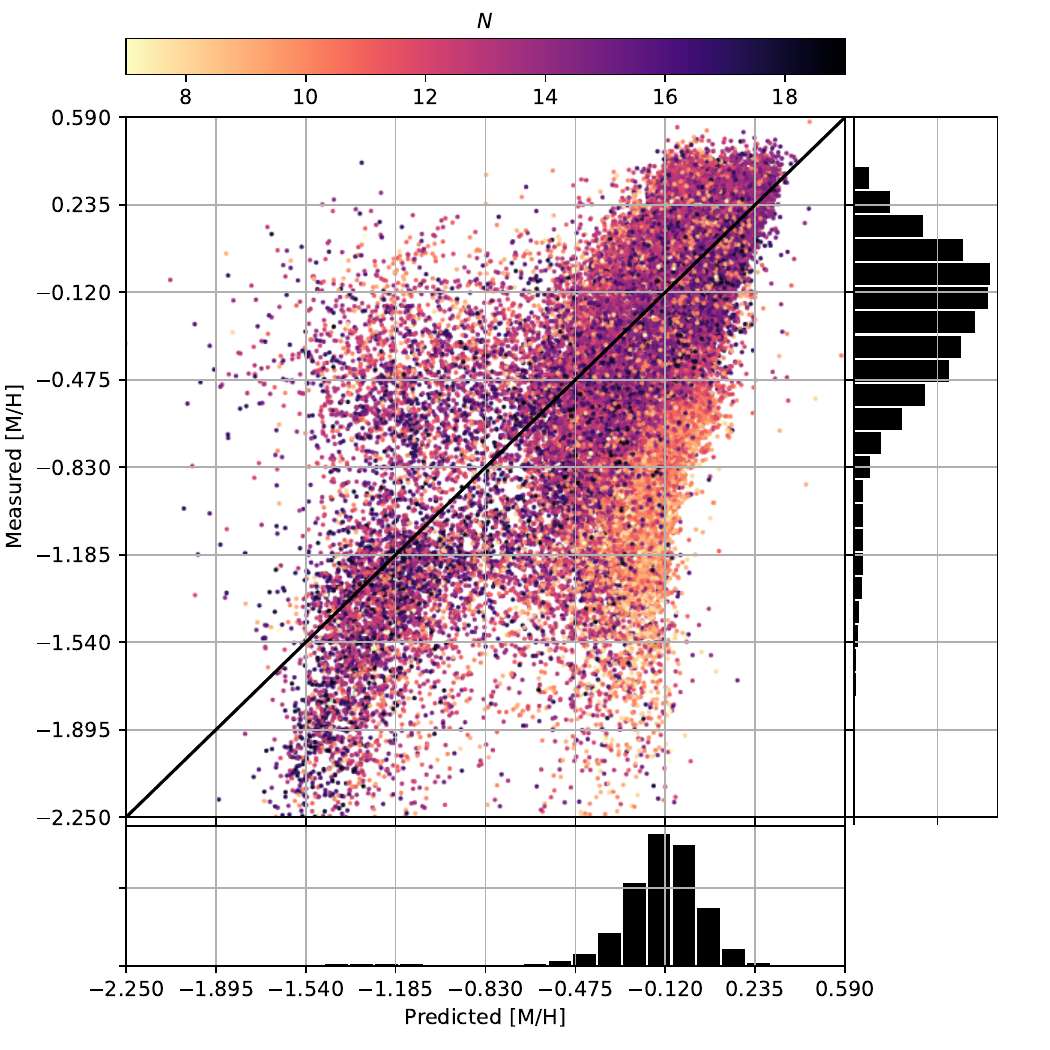}
    \caption{Predicted and measured metallicities for stars in the Halo sample. Color represents the number of measurments used to infer the stellar properties.}
    \label{fig:halo}
\end{figure}

For metallicity evaluation, we focus on stars in low-extinction regions ($b > 20^\circ$),
forming the \textit{Halo sample}. This parameter is notoriously difficult to estimate from
photometry alone, yet remains vital for numerous astrophysical investigations.
The APOGEE survey measures the metallicity with accuracy of the order of $\sim 0.05$ dex.
We select 129,066 stars from APOGEE, again ensuring each has at least 6 independent photometric observations
through cross-matching with external catalogs. The model's metallicity predictions,
taken as the median of 64 posterior samples, are then compared to APOGEE values.
The resulting performance yields an RMS error of $0.31$ dex and a MAD of $0.21$ dex.
The comparison between predicted and true metallicity is presented in Figure \ref{fig:halo}.

Sampling for all stars was completed in under 25 minutes on a single NVIDIA A100 GPU,
corresponding to a throughput of approximately $14 \cdot 10^3$ samples per second.
This represents a substantial speedup—by 3 to 4 orders of magnitude—over traditional CPU-based MCMC methods.

\begin{figure}
    \includegraphics[width = 0.6\textwidth]{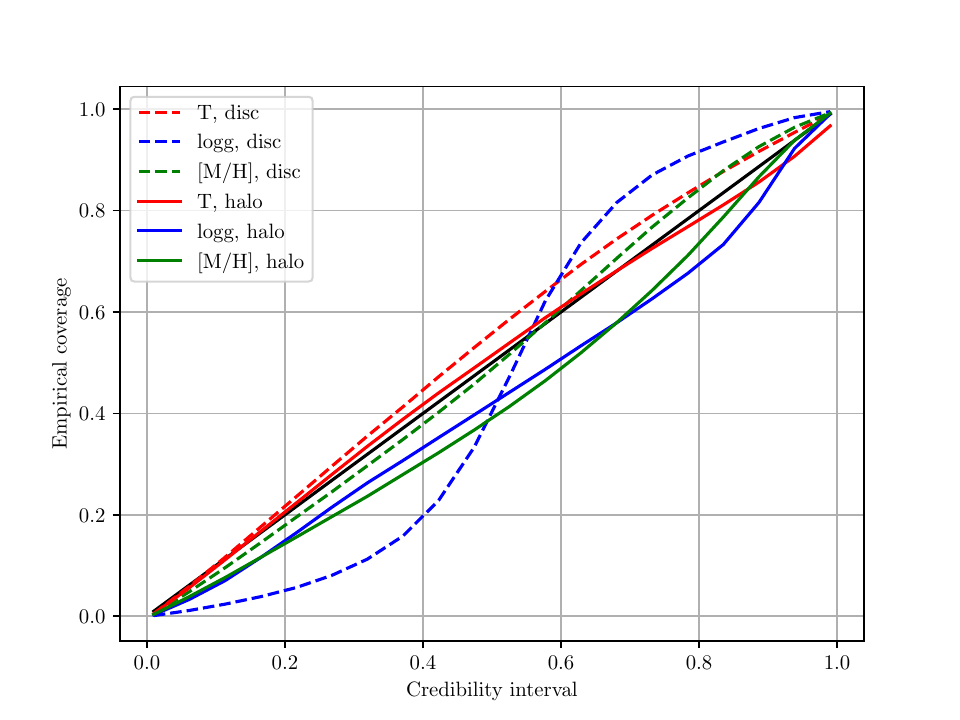}
    \caption{The coverage plot for the APOGEE sample with the distinction between the Disc and the Halo samples.}
    \label{fig:coverage}
\end{figure}

Tu further investigate the performance of the model, Bayesian coverage analysis was performed. To do so, we sampled the posterior distribution 
for all stars in the sample, creating HDI (Highest Density Interval) for each star and for various condifence intervals. HDI intervals were computed for 
each of the parameters observed by the APOGEE separately. Then, for different confidence ranges we checked how many true values lie in the HDI 
interval. Results are presented in Figure \ref{fig:coverage}. For parameters which correspond to lines
that are above the black line, the methods is underconfident and for those that fall below the black line, the method is overconfident. Here, we can 
see that the lines roughly follow the 
black line, indicating a close match with true Bayesian posterior.

\section{Conclusions}
We present a novel approach for fitting Spectral Energy Distribution (SED) models using Neural Posterior Estimation (NPE).
Our method enables seamless integration of photometric filters directly into the model,
providing flexible and efficient inference applicable to real-world scenarios.
Compared to traditional MCMC sampling performed on CPUs, our approach achieves several orders of magnitude speedup.

The model demonstrates excellent performance in estimating stellar temperatures, even for highly reddened sources.
As such, it is particularly well-suited for accelerating the construction of three-dimensional dust maps in the Galaxy.

However, estimating metallicity proves more challenging. This result is somewhat expected:
stars can exhibit significantly different metal abundances, often requiring distinct spectral templates.
Moreover, the presence of other error sources—particularly out-of-distribution (OOD) errors—further complicates the inference.

OOD errors are an inherent issue in large astronomical datasets. Skewed measurements due to mismatches,
photometric errors, or underestimated uncertainties can cause the model to fail.
These errors are difficult to detect and interpret. It is well-known that normalizing flows can
assign high probability densities to OOD inputs, exacerbating the issue \citep{OOD_flow}.

\bibliography{Iris-ML}
\bibliographystyle{icml2025}

\end{document}